\documentstyle[aas2pp4]{article}

\slugcomment{Accepted for publication on The Astrophysical Journal,
Part 1.}

\lefthead{M\'endez and Minniti}
\righthead{The nature of dark matter in the Galaxy}

\begin{document}

\title{Faint blue objects on the Hubble Deep Field North \& South as
possible nearby old halo white dwarfs}

\author{R. A. M\'endez}
\affil{Cerro Tololo Inter-American Observatory, Casilla 603, La
Serena, Chile. E-mail: rmendez@noao.edu}

\author{D. Minniti}
\affil{Pontificia Universidad Cat\'olica de Chile, Vicu\~na MacKenna
4860, Santiago, Chile. E-mail: dante@astro.puc.cl}

\altaffiltext{1}{Based on observations collected with the Hubble Space
Telescope. HST is operated by AURA Inc., under contract with the
National Science Foundation.}

\begin{abstract}

Using data derived from the deepest and finest angular resolution
images of the universe yet acquired by astronomers at optical
wavelengths using the Hubble Space Telescope (HST) in two
postage-stamp sections of the sky (Williams et al. 1996a,b), plus
simple geometrical and scaling arguments, we demonstrate that the
faint blue population of point-source objects detected on those two
fields (M\'endez et al. 1996) could actually be ancient halo white
dwarfs at distances closer than about 2~kpc from the Sun. This finding
has profound implications, as the mass density of the detected objects
would account for about half of the missing dark matter in the
Milky-Way (Bahcall and Soneira 1980), thus solving one of the most
controversial issues of modern astrophysics (Trimble 1987, Ashman
1992).

The existence of these faint blue objects points to a very large mass
locked into ancient halo white dwarfs. Our estimate indicates that
they could account for as much as half of the dark matter in our
Galaxy, confirming the suggestions of the MACHO microlensing
experiment (Alcock et al. 1997). Because of the importance of this
discovery, deep follow-up observations with HST within the next two
years would be needed to determine more accurately the kinematics
(tangential motions) for these faint blue old white dwarfs.

\end{abstract}


\section{Introduction}

The Hubble Deep Field data, in its Northern (HDF-N) and Southern
(HDF-S) versions, have been heavily used to study the evolution of
very distant galaxies (Livio, Fall and Madau 1998), back to times when
the Universe was only a fraction of its present age. Indeed, the main
motivation for the acquisition of these very deep images was the study
of a small portion of the sky to an unprecedented depth. Yet, these
data have also allowed astronomers to study and characterize the faint
objects that compose our own Galaxy. In a series of papers (Elson et
al. 1996, Flynn et al. 1996, M\'endez et al 1996), the stellar sample
derived from the HDF-N was used to set constraints on the faint-end of
the luminosity function for normal halo field stars, while the
shallower, but larger solid-angle data from the HDF-N flanking-fields
was used to better refine Galactic structural parameters (M\'endez and
Guzm\'an 1998).

In this paper we compare the stellar samples derived from the HDF-N \&
S pointings. Using a very simple, yet robust, model-independent
argument, we demonstrate that the faint blue-objects found on the
HDF-N (Elson et al. 1996, M\'endez et al. 1996) are indeed Galactic
stars, and not distant star-forming regions or compact galaxies as
previously suggested (Elson et al. 1996). Our finding is corroborated
by independent preliminary tangential motion measurements detected for
these faint blue stars (Ibata et al. 1998, Ibata 1998), which also
proves that they are not distant Galaxies, but rather nearby
stars. Recent evolutionary tracks (Hansen 1998) indicate that, if
these faint blue objects are Galactic, then they would be old halo
white dwarfs, with ages in the range 10-12~Gyr located at distances
of~1 to 2~kpc from the Sun.

\section{Point Sources in the Hubble Deep Fields North \& South}

We have used the source catalogues produced by the Space Telescope
Science Institute (STScI) from the combined (deepest), drizzled
(spatial resolution enhanced) images from HDF-N and HDF-S available
through the WWW. We note that the HDF-N catalogue used here is based
on a re-reduction of the HDF-N images, providing some 10\% increase in
depth or, correspondingly, better signal-to-noise for the brighter
sources, than that available from the original images used to detect
the faint blue objects. Both, the HDF-N and HDF-S catalogues have been
derived using exactly the same algorithms, procedures, and parameters,
and thus, they comprise a very homogeneous and self-consistent
dataset.

A critical step is the classification of point-sources {\it vs.}
extended objects. This has been achieved by using a widely tested
classifier trained with real images so as to provide the most robust
separation of stars {\it vs.}  extended objects by using a
neural-network scheme (Bertin, 1995). Figure~\ref{classdist} shows the
distribution of CLASS {\it vs.} magnitude for HDF N \& S. CLASS is the
probability that SExtractor assigns to an object as being point-like,
with CLASS=0 being an extended source and CLASS=1 being a point-like
object (CLASS is not a binary classifier but rather a continuous
variable that can take any value from zero to one). This figure
clearly indicates that there is reliable star-galaxy separation until
about $V + I \sim 29$, and that both data sets are quite homogeneous
and comparable in depth.

\section{The Faint blue Objects as White Dwarfs}

Figure~\ref{classhist} shows the frequency of CLASS as a function of
this parameter. The large peak at low values of CLASS indicate that
the sample is indeed dominated by galaxies, while the smaller, yet
conspicuous, peak at larger values of CLASS reveals the truly
point-like objects. Visual inspection reveals that all objects with
CLASS$< 0.85$ are clearly extended, and thus we have used the very
conservative cut at CLASS~$>0.90$ to select our stars. In addition,
the original source catalogues provided by STScI had to be trimmed to
avoid the many spurious detections near the detector boundaries were
the lower signal-to-noise leads to very high source confusion and
poor-photometry and shape classification. In the end, our sample
consists of 78 point-sources from HDF-N (solid angle of
4.334~arc-min$^2$) and 98 sources from HDF-S (solid angle of
4.062~arc-min$^2$). Photometry for these objects was calibrated using
the precepts described by M\'endez and Guzm\'an (1998). Worst-case
magnitude limits for the shallower HDF-S data (see
Figure~\ref{classdist} caption) have been computed from the STScI
exposure-time calculator available through their WWW pages. These
limits are used here only as a guide, and they do not have a critical
impact on the conclusions of this paper. Since the sample analyzed in
this paper is several times brighter than the magnitude limit, and
the field is not crowded, we do not have to apply any completeness
corrections, which should be negligible above 5$\sigma$ the sky level,
as shown by Paresce et al. (1996) on deep HST images of the globular
cluster NGC~6397, acquired with the WFPC2. Actually, our discussion is
restricted here to those sources above 15$\sigma$ the sky level (see
Figure~\ref{figcmd1}.)

Figure~\ref{figcmd1} shows the calibrated color-magnitude diagrams
derived from the catalogues. The faint blue stars are clearly seen in
both figures, but they appear more numerous on the HDF-S
sample. Indeed, this simple fact provides the central argument of this
paper. HDF-N is located at Galactic coordinates
$(l,b)=(125.89^o,+54.83^o)$, while HDF-S is located at
$(l,b)=(328.25^o,-49.21^o)$. Therefore, HDF-N is looking towards the
outer portion of the Milky-Way, while HDF-S looks inward. It is well
known that the stellar density decreases as a function of distance
from the Galactic center, either in an exponential fashion for
disk-like stars, or as a power of the distance for halo stars
(Majewski 1993). Therefore, one would naturally expect to see more
stars towards the HDF-S than towards the HDF-N. Is this actually the
case?  Figure~\ref{figcmd1} shows the locus for M-dwarfs belonging to
the Galactic disk at Heliocentric distances of 1~kpc, and M-subdwarfs
belonging to the Galactic halo at distances of 8~kpc, derived from the
best available trigonometric parallaxes for these two types of stars
(Monet et al. 1992). The characteristic distances adopted for these
two types of stars correspond to the typical distances that one
expects for them at these magnitudes and Galactic position, as derived
from a Galactic model which reproduces the observed HDF-N and Flanking
Fields magnitude- and color-counts (M\'endez et al. 1996, M\'endez and
Guzm\'an 1998).

From Figure~\ref{figcmd1} (see also Table~1) we see that on HDF-N
there are 10 stars within the boundaries allowed by the M-dwarf and
subdwarf sequences, while the number of similar objects on HDF-S is
22. Their ratio is roughly a factor of 2. Whether the absolute numbers
observed in each field within those boundaries are what one would
expect from a standard Galactic model is actually irrelevant to this
discussion (however, it has been already shown that the Galactic model
predictions and the observed counts do agree on HDF-N and its flanking
fields (M\'endez et al. 1996, M\'endez and Guzm\'an 1998)). If the
faint blue objects are actually extragalactic in nature, as it has
been suggested (Elson et al. 1996), then one should not see a
variation in their numbers when going from HDF-N to HDF-S, since the
Universe is isotropic on large scales (see Table~1). We should remind
the reader that, if these objects are assumed to be extragalactic,
they would be located at redshifts of $z \ge 1$ (Elson et al. 1996),
and at these scales the angular correlation function (which measures
the number of galaxy pairs at a given angular separation, in excess of
a random distribution) has been found to be zero to within $5 \times
10^{-4}$ for angular separations larger than 6$^o$ (Maddox et
al. 1990, the HDF-N and HDF-S are 165$^o$ apart in the sky). However,
what we find from Figure~\ref{figcmd1} is that the number of faint
blue objects is 5 on HDF-N and 10 on HDF-S, their ratio being a factor
of two, almost exactly as it is for normal stars. This argument
suggests that the faint blue objects are not extragalactic and that,
furthermore, their space distribution follows that of normal Galactic
stars. Assuming a Poisson distribution, the probability of seeing 10
sources on HDF-S when 5 are expected is only 1.3\% (account has to be
made for the different solid-angles covered by both
samples). Therefore, even though the samples are small, the observed
difference in the number of expected objects if they had an isotropic
N-S distribution is highly significant. Additional indication that the
faint blue objects are actually nearby is provided by Ibata and Lewis
(1998, and Ibata 1998) who have obtained preliminary tangential
motions for the point-sources on HDF-N using a two-year baseline on
HST. They find that four of the five faint blue objects do have
detectable tangential motions at a 3$\sigma$ level or more, thus
ruling out the hypothesis that they are extragalactic objects (their
motions are actually consistent with halo kinematics at 1~to 2~kpc,
see below). A more robust proper-motion determination would require
additional observations with HST within the next two years to increase
the time baseline for the tangential motion measurement.

\section{Conclusions and implications for the nature of dark matter in
the Galaxy}

The MACHO microlensing experiment has found that a significant
fraction of the dark matter is baryonic, and made of objects with
0.5~$M_\odot$ (Alcock et al. 1997). They have suggested white dwarfs
(WDs) as possible candidates because they have the right mass and,
though very numerous, old ones would be quite faint to have remained
undetected thus far. Figure~\ref{figcmd1} shows the locus of old WDs
as recently computed by Hansen (1998) using the latest atmospheric
models and opacity tables, and confirmed observationally on the cool
\& low-luminosity WD LHS~3250 (Harris et al. 1999). It is clear from
this figure that the faint blue objects do fall in the region
predicted by these models, as originally pointed out by Hansen himself
for HDF-N. This fact, plus the discussion in the preceding paragraph,
indicates that we have detected a population of old faint and blue
white dwarfs belonging to the Galactic halo, and located at
Heliocentric distances of up to 2~kpc. This finding has profound
implications for the nature of dark matter in our own galaxy. If the
objects that we have detected are actually old WDs from the halo, then
their expected number in the HDF-N {\it vs.} the HDF-S should follow
that of the general halo population. Is this actually case? For a
density law similar to that exhibited by halo field tracers (e.g.,
RR-Lyraes or Blue Horizontal-Branch stars, (Sluis and Arnold 1998)),
i.e. $\rho \sim R^{-3}$ with an axial ratio of 0.8, where R is the
distance from the Galactic center, we find that the expected number
ratio of halo stars between the HDF-N and HDF-S is about
1.71. Therefore, the factor of two increase when going from HDF-N to
HDF-S is consistent with their being associated with the halo field
(given the uncertainties in the halo density law).

If, as the preceding discussion suggests, we have detected a
population of faint blue old halo WDs in the vicinity of the Sun. What
is their contribution to the local Galactic mass budget? Their mass
contribution ($M_{\rm Halo \, WD}$) is given by:

\begin{equation}
M_{\rm Halo \, WD} = N_{\rm Halo \, WD} \times \mu_{\rm WD} = 8.46
\times 10^{-8} \, \Omega \, \rho_\odot \, R_\odot^{3}
\int_{0}^{d_{lim}} \frac{r^2}{R^{3}} dr
\end{equation}

where $N_{\rm Halo \, WD}$ is the number of halo WDs with typical mass
$\mu_{\rm WD}$ observed on HDF, $\Omega$ is the solid angle (in
squared arc-min) subtended by the HDF, $\rho_\odot$ is the mass
density of this component in the solar neighborhood (in $M_\odot /
pc^3$), $R_\odot$ is the solar Galactocentric distance (in pc), $r$ is
the Heliocentric distance, and $d_{lim}$ is the maximum Heliocentric
distance sampled by the HDF data.

Assuming a typical WD mass of 0.6~$ M_\odot$, and a maximum sampling
distance of 2~kpc (see Figure~\ref{figcmd1}), we obtain a value of
$4.64^{+0.33}_{-0.66} \times 10^{-3}$~$M_\odot / pc^3$ for the local
($R_\odot = 7.5$~kpc) mass density of halo WD. On the other hand,
dynamical studies of the Galaxy indicate a {\it local} value for the
mass density of dark matter of $1.26 \times 10^{-2}$~$M_\odot / pc^3$
(this mass density is equivalent to 0.19~$M_\odot / pc^3$ at the
Galactic center on a density law of the type adopted by Bahcall and
Soneira 1980). Therefore, our HDF sample accounts for about 1/3 to 1/2
of the dark matter in the Milky Way.

The analysis of gravitational microlensing events of stars in the
Large Magellanic Clouds (Alcock et al. 1997) places the mass of the
lensing objects in the range 0.5$^{+0.3}_{-0.2}$~$M_\odot$, suggesting
that they might actually be old WDs. Furthermore, the MACHO
collaboration finds that about half of the dark matter halo of the
Milky Way could be composed of those old WDs. These two suggestions by
the MACHO group seem to be in agreement with our analysis of the HDF
faint blue data and, basically, resolves about half of the dark matter
problem in our Galaxy (the other half still being unaccounted for by
ordinary matter).

Our derived mass density assumes that {\it all} of the faint blue
objects on both HDF-N \& S are indeed halo white dwarfs. However, it
seems that at least one of these objects is quite close to the halo
subdwarf sequence on HDF-N (see Fig.~(\ref{figcmd1}), upper panel),
while some of the fainter stars on HDF-S (see Fig.~(\ref{figcmd1}),
lower panel) have colors that are not inconsistent with those of M
subdwarfs, although being some 2 magnitude fainter in I than normal M
subdwarfs. As mentioned earlier, M\'endez et al. (1996) have
demonstrated that the faint blue population could not be accounted for
with current Galactic models, and therefore required something
new. Since no similar comparison has been produced yet for HDF-S, we
have run some Galactic models to predict the number and distribution
of M-dwarfs from the disk and halo from the same model used for the
analysis of the HDF-N. In the range $20 \le V \le 26$ and $0 \le B-V
\le 1.9$, mostly encompassing the locus of stars on
Fig.~\ref{figcmd1}, the model described by M\'endez et al. (1996)
(with the changes indicated by M\'endez and Guzm\'an 1998) predicts
between 18 and 22 stars, depending on one's choice of scale-heigh for
main-sequence disk stars (still a somewhat uncertain model
parameter). This good agreement of the normal M dwarfs with the models
(see Table~1) is, of course, no compelling assurance for the ``need''
of a new population of stars, since the samples are still quite small:
A $1 \sigma$ Poisson fluctuation on the count of 22 model-predicted
stars would account for a large fraction (almost 5 out of 10) of the
faint blue stars as being part of the stellar content of normal models
- which do not include halo WDs.

We should also consider that errors intrinsic to the calculation of WD
tracks are actually not large, 0.1 mags or less, and mostly coming
from uncertainties in the temperature and pressure behavior in the
atmosphere (cool white dwarfs are convective to the photosphere),
although the results seem to be fairly insensitive to mixing length
prescriptions (Hansen 1999). Another uncertainty is from opacities not
as yet included in the models, in particular the higher $H_2$ level
transitions. Even though lower transitions are dominant, these higher
transitions might take notches out of the flux between the broad
absorption bands, and this would likely make WD appear bluer. Given
that the lower bands are the dominant opacity contributor, the general
trend should be robust, but the detailed colors could vary somewhat,
perhaps as much as 0.5 magnitudes. Finally, a bigger uncertainty,
which can reach several tenths of a magnitude, comes from the fact
that cool WD atmospheres are distinctly non-blackbody and have many
spectral signatures, making them quite susceptible to the adopted
bandpasses. Calculations for both HST and Johnson \& Kron-Cousins
colors can differ by up to about 0.5 mags. Hansen has adopted, for the
models used here, the synthetic magnitudes described in Section~5.2 of
Holtzman et al. (1995). This was done because transformations such as
those described in the earlier sections of the Holtzmann paper are
based on standard stars, and will not hold well for stars whose
spectra do not resemble the standard stars. Furthermore, the procedure
in section 5.2 also matches well with those used in other WD
atmosphere calculations, making comparisons easier. After calculating
the WFPC2 magnitudes, magnitudes and colors on the VRI system were
computed using the transformations defined on Table~10 of the Holtzman
et al. paper, following thus the same procedure employed to convert
our observed HST magnitudes to the Johnson-Cousins system. Given all
these uncertainties, it is reassuring that the more recent independent
model calculations by Saumon and Jacobson (1999), and which overcome
some of the simplifying assumptions of the earlier models, do agree
with Hansen's original calculations. In particular, Saumon and
Jacobson conclude that very cool ($T_{\rm eff} < 3500$~K) halo
($t_{age} > 10$~Gyr) WDs would have V-I~$< 1.4$.

\acknowledgments

R.A.M acknowledges Drs. W. Couch, T.M. Girard and S. Zepf for their
help, to Drs. R. Ibata and B. Hansen for information prior to
publication, and to CTIO and the RGF foundation for their generous
support during his stay at Yale. D.M. acknowledges the support of a
Chilean Fondecyt grant N$^o$ 01990440 and DIPUC, and the
U.S. Department of Energy through Contract W-7405-Eng-48 to the
Lawrence Livermore National Laboratory.

\clearpage

\begin{deluxetable}{ccccc}

\tablecaption{
Observed number of Galactic stars, faint blue point sources, and
extragalactic objects in the HDF-S and HDF-N.
\label{tab1}}

\tablewidth{0pt}
\tablehead{
\colhead{Object} &
\colhead{HDF - N} &
\colhead{HDF - N (norm)\tablenotemark{a}} &
\colhead{HDF - S} &
\colhead{Normalized ratio HDF-S / HDF-N\tablenotemark{b}}
}
\startdata
Galactic stars & 10  &   9.37 &  22 & $2.35 \pm 0.90$ \nl
Faint blue     & 5   &   4.69 &  10 & $2.13 \pm 1.17$ \nl
Extragalactic\tablenotemark{c}  & 566 & 530.48 & 486 & $0.916 \pm 0.057$ \nl
\tablenotetext{a}{Normalized to same solid angle as HDF-S.}
\tablenotetext{b}{Poisson noise from the original, unnormalized,
counts.}
\tablenotetext{c}{Galaxies selected in same magnitude and color range
as the faint blue sources.}
\enddata
\end{deluxetable}

\clearpage

\figcaption[class_hdf.ps]{SExtractor CLASS parameter {\it vs.} HST V+I
(uncalibrated) magnitude from the co-added (and deepest) F606W and
F814W drizzled-combined frames produced by STScI for HDF-N (upper
panel) and HDF-S (lower panel). The total effective on-target
integration times that went into the combined V+I frames are 64.63 and
50.44 hours for the northern and southern deep fields,
respectively. It is apparent that we can reliably separate
point-sources (CLASS~$\sim 1$) from extended objects (CLASS~$\sim 0$)
down to $V + I \sim 29$.
\label{classdist}}

\figcaption[class_histo_hdf.ps]{Specific frequency of stars as a
function of CLASS for HDF-N (blue) and HDF-S (red). We have trimmed
our point-like sample in a very conservative way at CLASS~$\ge
0.90$. Objects with CLASS$< 0.85$ are clearly extended in the
individual V and I drizzled-combined frames. Objects with $0.85
\le$~CLASS~$< 0.90$ fall below our magnitude cutoff (see
Figure~\ref{figcmd1}), and therefore do not affect the conclusions of
this paper (see text).
\label{classhist}}

\figcaption[cmd_color_ns3.ps]{Color-magnitude diagrams in calibrated
Johnson-Cousins I {\it vs.} V-I for point sources from the HDF-N
(upper panel) and the HDF-S (lower panel). The red dotted line is the
locus for disk M-dwarfs at 1~kpc from the Sun, while the green dotted
line is the locus for sub-dwarfs at a Heliocentric distance of
8~kpc. The solid black lines indicate the 15$\sigma$ magnitude limits
imposed by the exposure time in the combined HDF I frame (horizontal
line) and the V combined frame (diagonal line). Only objects above the
intersection of these two solid lines are firm detections, with good
star-galaxy separation. The red stars are bona-fide Galactic stars as
predicted, in number and location on the CMD, from standard Galactic
models. The faint blue stars are shown by the blue symbol. The true
nature of these objects is revealed as old, cold halo WDs (see
text). The blue dotted line indicates the theoretical predicted locus
for halo WDs of 0.6~$M_\odot$ at 1~kpc, while the solid line indicates
the locus for the same stars at 2~kpc. WDs in the mass range 0.5 to
0.9~$M_\odot$, encompassing the full range of models computed by
Hansen, exhibit a similar color magnitude distribution. The bluening
of the WD tracks occurs at an age of about 10~Gyr and $T_{\rm eff}
\sim 3,500$~K. Uncertainties in the physics of the models and the
transformation of its predictions to the observational plane can
account for up to 0.5~mag in the predicted WD colors (Hansen 1999).
\label{figcmd1}}

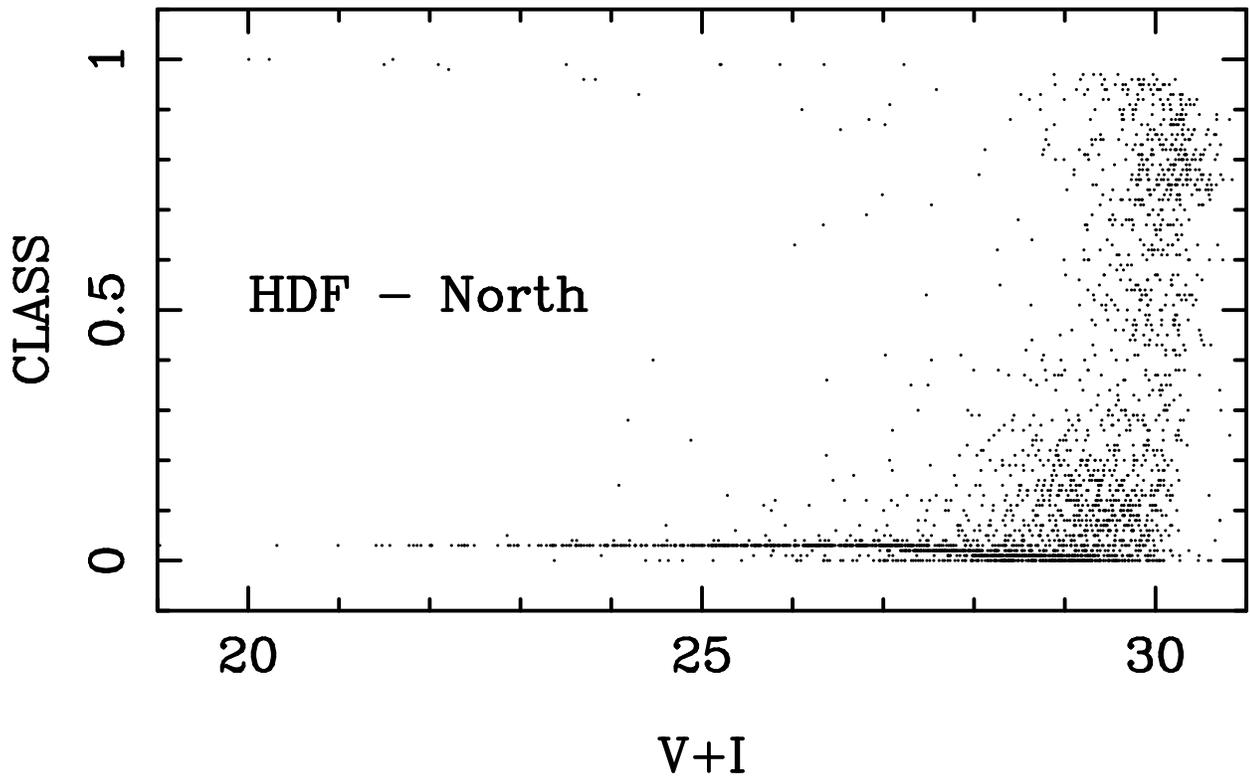
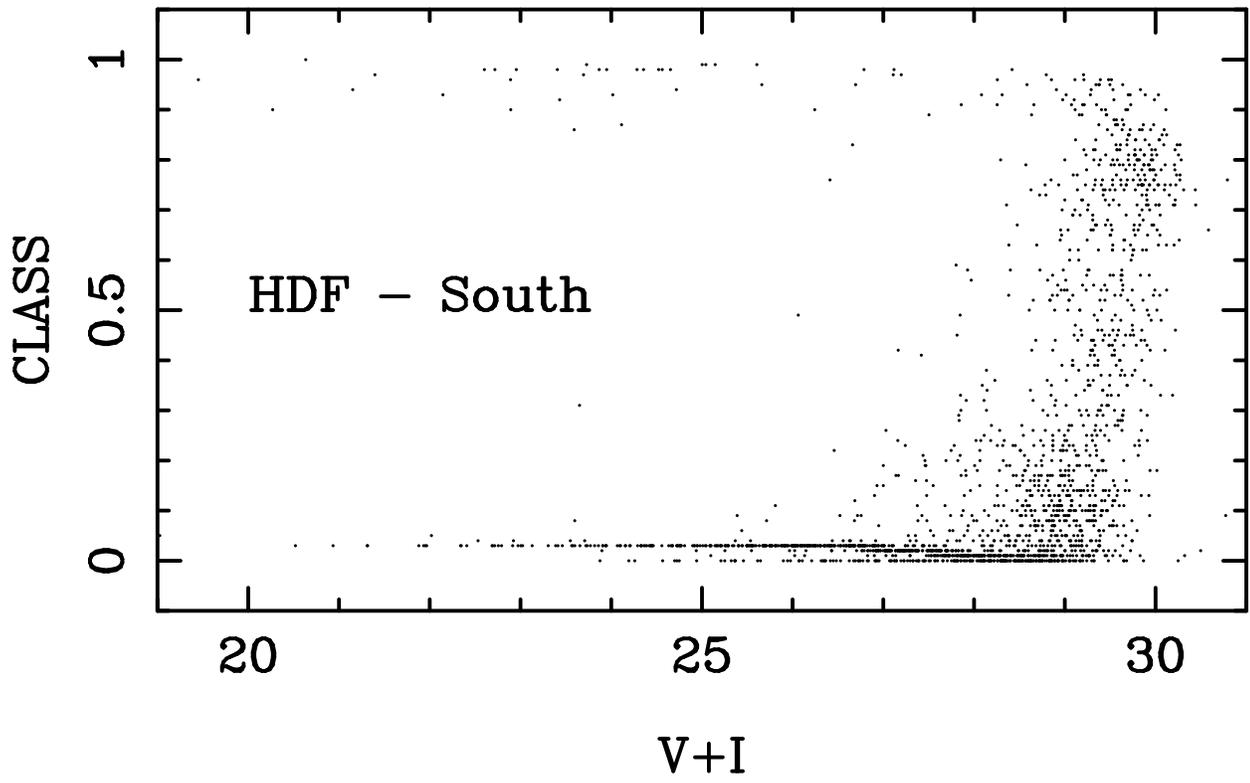
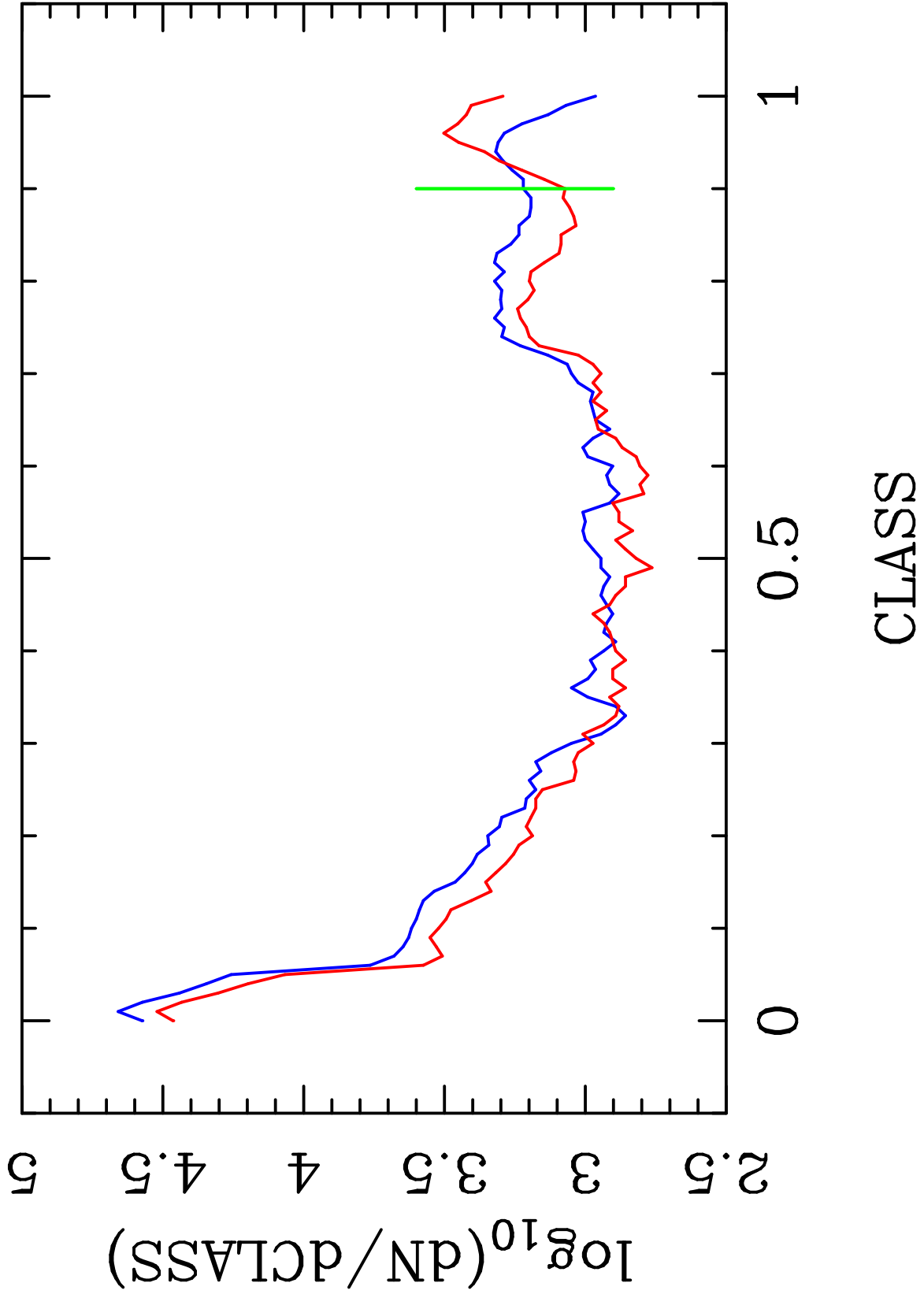
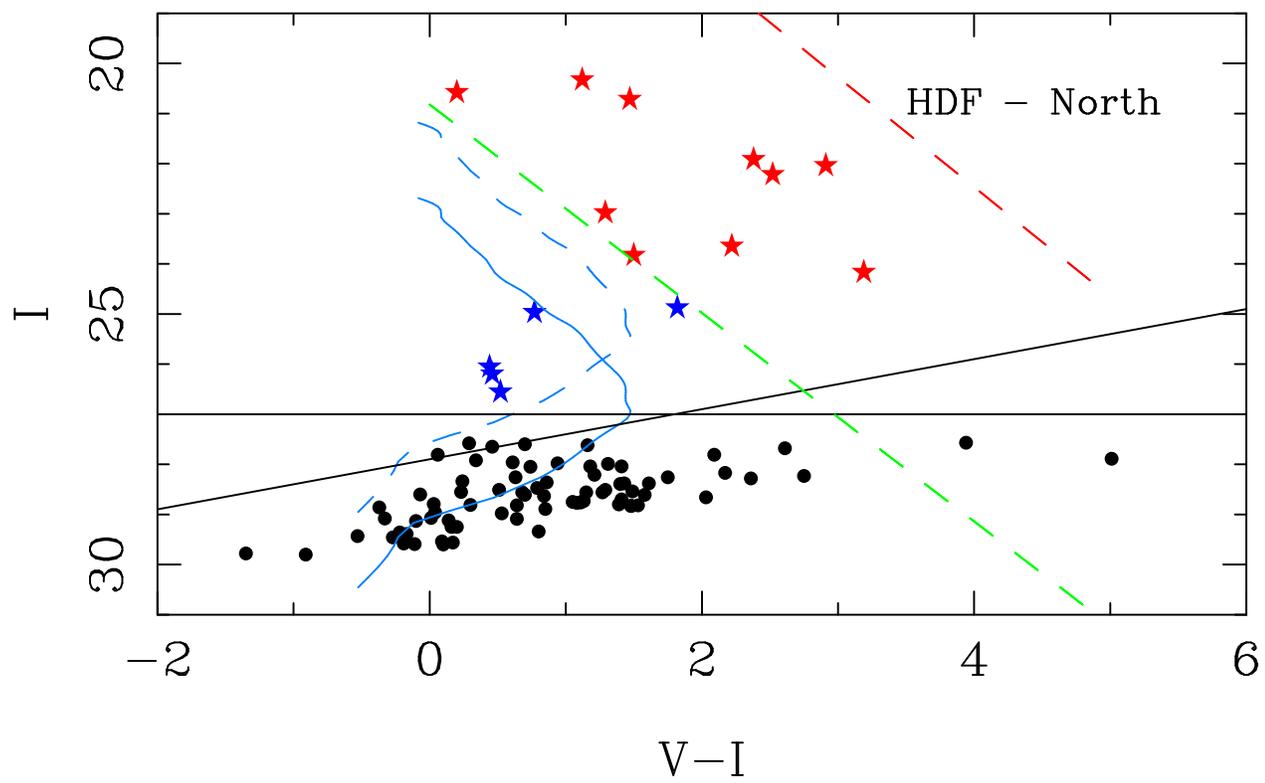
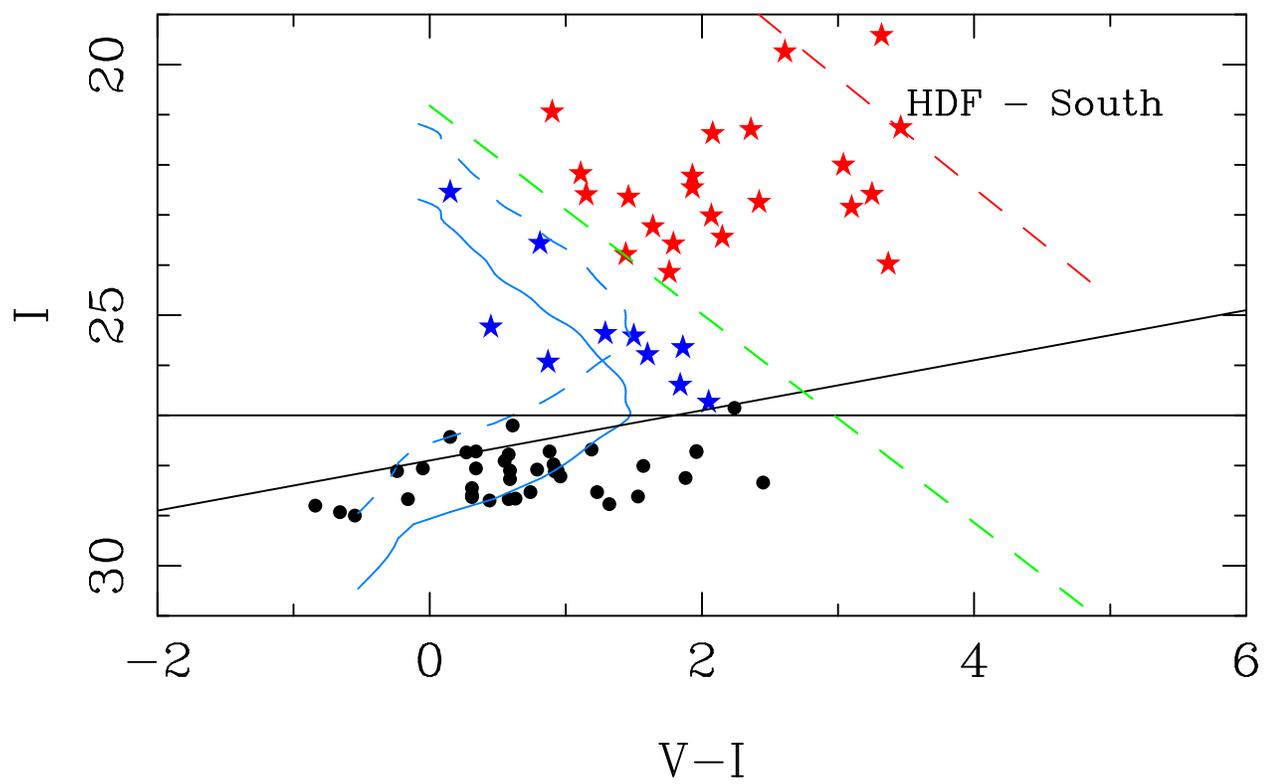
\end{document}